# X-ray Scaling Relations of 'core' and 'coreless' E and S0 Galaxies

**Dong-Woo Kim and Giuseppina Fabbiano**

Smithsonian Astrophysical Observatory,
60 Garden Street, Cambridge, MA 02138, USA

(July 22, 2015)

## Abstract

We have re-examined the two X-ray scaling relations of early-type galaxies (ETGs), $L_{X,GAS} - L_K$ and $L_{X,GAS} - T_{GAS}$, using 61 ATLAS$^{3D}$ E and S0 galaxies observed with *Chandra* (including ROSAT results for a few X-ray bright galaxies with extended hot gas). With this sample, which doubles the number of ETGs available for study, we confirm the strong, steep correlations reported by Boroson et al (2011). Moreover, the larger sample allows us to investigate the effect of structural and dynamical properties of ETGs in these relations. Using the sub-sample of 11 'genuine' E galaxies with central surface brightness cores, slow stellar rotations and old stellar populations, we find that the scatter of the correlations is strongly reduced, yielding an extremely tight relation, of the form $L_{X,GAS} \sim T_{GAS}^{4.5\pm0.3}$. The rms deviation is only 0.13 dex. For the gas-rich galaxies in this sample ($L_{X,GAS} > 10^{40}$ erg s$^{-1}$), this relation is consistent with recent simulations for velocity dispersion supported E galaxies. However, the tight $L_{X,GAS}$-$T_{GAS}$ relation of genuine E galaxies extends down into the $L_X \sim 10^{38}$ erg s$^{-1}$ range, where simulations predict the gas to be in outflow/wind state, with resulting $L_X$ much lower than observed in our sample. The observed correlation may suggest the presence of small bound hot halos even in this low luminosity range. At the high luminosity end, the $L_{X,GAS}$ - $T_{GAS}$ correlation of core elliptical galaxies is similar to that found in samples of cD galaxies and groups, but shifted down toward relatively lower $L_{X,GAS}$ for a given $T_{GAS}$. In particular cDs, central dominant galaxies sitting at the bottom of the potential well imposed by the group dark matter, have an order of magnitude higher $L_{X,GAS}$ than our sample core galaxies for the same $L_K$ and $T_{GAS}$. We suggest that enhanced cooling in cDs, which have higher hot gas densities and lower entropies, could lower $T_{GAS}$ to the range observed in giant Es; this conclusion is supported by the presence of extended cold gas in several cDs. Instead, in the sub-sample of coreless ETGs – these galaxies also tend to show stellar rotation, a flattened galaxy figure and rejuvenation of the stellar population – $L_{X,GAS}$ and $T_{GAS}$ are not correlated. The $L_X$-$T_{GAS}$ distribution of coreless ETGs is a scatter diagram clustered at $L_{X,GAS} < 10^{40}$ erg s$^{-1}$, similar to that reported for the hot interstellar medium (ISM) of spiral galaxies, suggesting that both the energy input from star formation and the effect of galactic rotation and flattening may disrupt the hot ISM.

# 1. INTRODUCTION

Two families of ETGs have been identified, based on the central surface brightness distribution: (1) core elliptical galaxies, which typically have large stellar masses (of which $L_K$ is a proxy), high stellar velocity dispersion ($\sigma$), round isophotes, uniformly old stellar populations with no sign of recent star formation (SF), and slowly rotating stellar kinematics; and (2) power-law (also called coreless) galaxies, which instead tend to be smaller, disky, rotating, and with some recent star formation. Core, $\sigma$-supported, genuine Es tend to have a larger amount of hot X-ray emitting gas (see Kormendy et al. 2009; Lauer 2012; Pellegrini 2005; Sarzi et al. 2013). Kormendy et al. (2009) suggested that core Es are primarily formed by dry mergers. The hot ISM of core Es may provide the working surface necessary for AGN feedback by storing and smoothing episodic energy input, and by shielding against the accretion of fresh gas, thus impeding star formation (Binney 2004; Nipoti & Binney 2007; Kormendy et al 2009; Gabor and Dave' 2014). Coreless galaxies, instead, may be the product of gas-rich wet mergers, with ensuing star formation (Kormendy 2009).

The X-ray properties of ETGs are important for constraining these scenarios. X-ray scaling relations have been widely used to investigate the origin and evolution of the hot ISM of ETGs (e.g., Fabbiano 1989; Mathews & Brighenti 2003; BKF; Kim & Fabbiano 2013; Civano et al. 2014). Recent numerical simulation studies (e.g., Choi et al. 2014; Negri et al. 2014) have attempted to reproduce the observed scaling relations, to constrain the physical mechanisms that shape the hot ISM. While the correlation between optical and X-ray luminosities has been extensively investigated, most pre-*Chandra* studies have used the total X-ray luminosity ($L_{X,Total}$) which is close to the X-ray emission of the hot ISM ($L_{X,GAS}$) only for gas-rich X-ray luminous ETGs. *Chandra* observations have given us the ability to separate the different X-ray emission components of ETGs, extracting accurate measurements of $L_{X,GAS}$ (Boroson, Kim & Fabbiano 2011, hereafter BKF). This work has shown that the well-known factor of 100 spread (e.g., Fabbiano 1989) in the most studied ETG scaling relation, $L_X - L_{Optical}$, increases to a factor of 1000 when $L_{X,GAS}$ is used (BKF; Kim & Fabbiano 2013).

With *Chandra* we have also been able to measure the temperatures of these gaseous components, and derive the $L_{X,GAS}$ - $T_{GAS}$ scaling relation for ETGs. This relation is tighter than the $L_{X,GAS}$ - $L_K$ (BKF), at least for gas-rich galaxies ($L_{X,GAS} > 10^{40}$ erg s$^{-1}$), and may indicate virialization of the hot gas in the dark matter halos (Kim and Fabbiano 2013). The $L_{X,GAS}$ - $T_{GAS}$ relation of normal ETGs, however, differs in both normalization and slope from that observed in more massive systems: cluster, groups and cD galaxies (BKF; Pratt et al. 2009; O'Sullivan et al. 2003). Because the relative importance of baryonic physics over the gravitational pull of dark matter may be the cause of these differences, comparing scaling relations may constrain the processes effective in different mass ranges. For gas-poor ETGs ($L_{X,GAS} < 10^{40}$ erg s$^{-1}$), simulations suggest that the hot ISM is in the outflow/wind state, leading to larger $T_{GAS}$ and lower $L_{X,GAS}$. Moreover, galaxy rotation, flattened potential, embedded disks and recent star formation may affect the evolution of this ISM, and change its physical properties by altering the potential field and energy budget (e.g., Pellegrini 2011, Negri et al. 2014).

Recent detailed studies of the structural and dynamical properties of increasing larger ETG samples provide the baseline information to compare with the properties of the hot ISM. In this paper, we use the joint information of the ATLAS$^{3D}$ sample (Cappellari et al. 2011) and of *Chandra*, to take a new look at the X-ray scaling properties of 61 ETG. This is the largest sample that can be investigated to date with a uniform set of well-characterized properties. In Section 2, we describe our sample selection; in Section 3 we describe the *Chandra* observations and data reduction techniques; in Section 4, we present two X-ray scaling relations of ETGs, $L_{X,GAS} - L_K$ and $L_{X,GAS} - T_{GAS}$, respectively. In Section 5, we discuss the implications of our results and in Section 6, we summarize our results.

## 2. GALAXY SAMPLE

We use the ATLAS$^{3D}$ sample (Cappellari et al. 2011) for our study. ATLAS$^{3D}$ provides extensive multi-wavelength data and uniformly-derived galaxy properties, including: stellar rotation (Emsellem et al. 2011), central radial profile (Krajnovic et al. 2013), cold gas mass (CO, HI) and dust presence (references in Table 1). We select for our study the 61 ATLAS$^{3D}$ ETGs, which were observed with *Chandra* for an exposure time longer than 10 ksec. This joint ATLAS$^{3D}$ – *Chandra* selection doubles the size of the ETG sample previously studied in X-rays (30 galaxies, BKF), while providing a better characterization of the galaxy properties. The new sample includes 5 cD galaxies; cD galaxies were not included in the BKF sample.

In Table 1, we list the sample with information taken from a series of papers published by the ATLAS$^{3D}$ team (see Footnotes in Table 1.) Comparing the ATLAS$^{3D}$ properties with results published previously in the literature, we find generally good agreement for morphological type, central stellar velocity dispersion, central radial profile characteristics (core vs. coreless), and ellipticity. There are some discrepant measurements for a few galaxies, which we discuss below. These discrepancies do not change our results noticeably.

Since stellar ages are not included in the ATLAS$^{3D}$ catalog, we used a range of measurements available in the literature (see Table 1). These stellar ages are primarily luminosity-weighted averages, using single stellar population models. Therefore, they may be inaccurate in some galaxies where a small population of recently formed stars could mimic an overall young stellar population when weighted by luminosity (e.g., Thomas et al. 2005). Additionally, we use the masses of cold atomic and molecular gas, and the presence of dust (from the ATLAS$^{3D}$ publications – see Table 1), as an indirect indication of recent star formation.

Of the ATLAS$^{3D}$ galaxy properties, the central structure (core vs. coreless) is useful to separate two classes of ETGs (Kormendy et al. 2009; Lauer 2012). The morphological type could be misclassified due to hidden disks or dust obscuration. Classifying ETGs by other quantities ($\sigma$, age, ellipticity, rotation) is somewhat arbitrary because of no unambiguous dividing value. We take the classifications of core and coreless galaxies from Krajnovic et al (2013) who fit the Nuker law (Lauer et al. 1995). As extensively discussed in Krajnovic et al. (2013), different methods (e.g., core-Sersic model) may result in different classifications (see below for a few examples, see also Dullo & Graham 2013 for related issues). Although the Sersic core and Nuker core may not be structurally the same, there is no significant difference if one is only interested to separate cores from the rest of the profiles (see Appendix A in Krajnovic et al. 2013). We also apply stricter criteria to identify genuine ellipticals as core, passively evolving (an old stellar system with no sign of recent star formation), $\sigma$-supported (slow or no rotation), morphologically elliptical galaxies (in section 4.2 and 5.1).

Using 2D integral field spectroscopy, Emsellem et al. (2011) separated ETGs into two classes, fast and slow rotating galaxies. Instead of the conventional measure of stellar rotation (V/$\sigma$), they applied a new rotation parameter, $\lambda_R = <RV> / [ R (V^2+\sigma^2)^{0.5}]$ and classified as fast and slow rotators ETGs with $\lambda_R$ greater or smaller than 0.31 $\varepsilon^{0.5}$, respectively. Comparing the central structure and the stellar kinematics, Lauer (2012) suggested that the slow and fast rotating ETGs are essentially the same as the core and coreless galaxies, respectively, if a slightly different boundary ($\lambda_R = 0.25$) between slow and fast rotators is applied. In contrast, Krajnovic et al. (2013) suggested the presence of genuine populations of slowly rotating coreless galaxies and fast rotating core galaxies. In Figure 1, we plot our sample in the $\lambda_R$ - $\varepsilon$ plane. We identify core and coreless galaxies (as determined by Kranjnovic et al. 2013) with red filled squares and blue filled circles, respectively; galaxies with intermediate profile, with large blue open circles; galaxies with no measurement of the central radial profile with small blue open circles. The solid green curve is $\lambda_R = 0.31\ \varepsilon^{0.5}$ ( Emsellem et al. 2011; Krajnovic et al. 2013) and the dashed line is $\lambda_R = 0.25$ (Lauer 2012). Whatever the classification, on average core galaxies are slower rotators than coreless galaxies. In particular, fast rotating core galaxies (red squares above the green curve in Fig. 1), are the slowest among fast rotators. If we apply a cut at $\lambda_R = 0.2$ (slightly lower than that used by Lauer) to our sample, we would identify 17 out of 19 core galaxies as slow rotators with only 2 exceptions and 27 out of 29 coreless galaxies (or 22 out of 23 without intermediate ones) as fast rotators again with only 2 exceptions (or 1 exception without intermediate ones). Given the location in the $\lambda_R$ - $\varepsilon$ plane, the 13 galaxies with unknown central structure (small open circles in Figure 1) are likely to be fast rotating coreless galaxies.

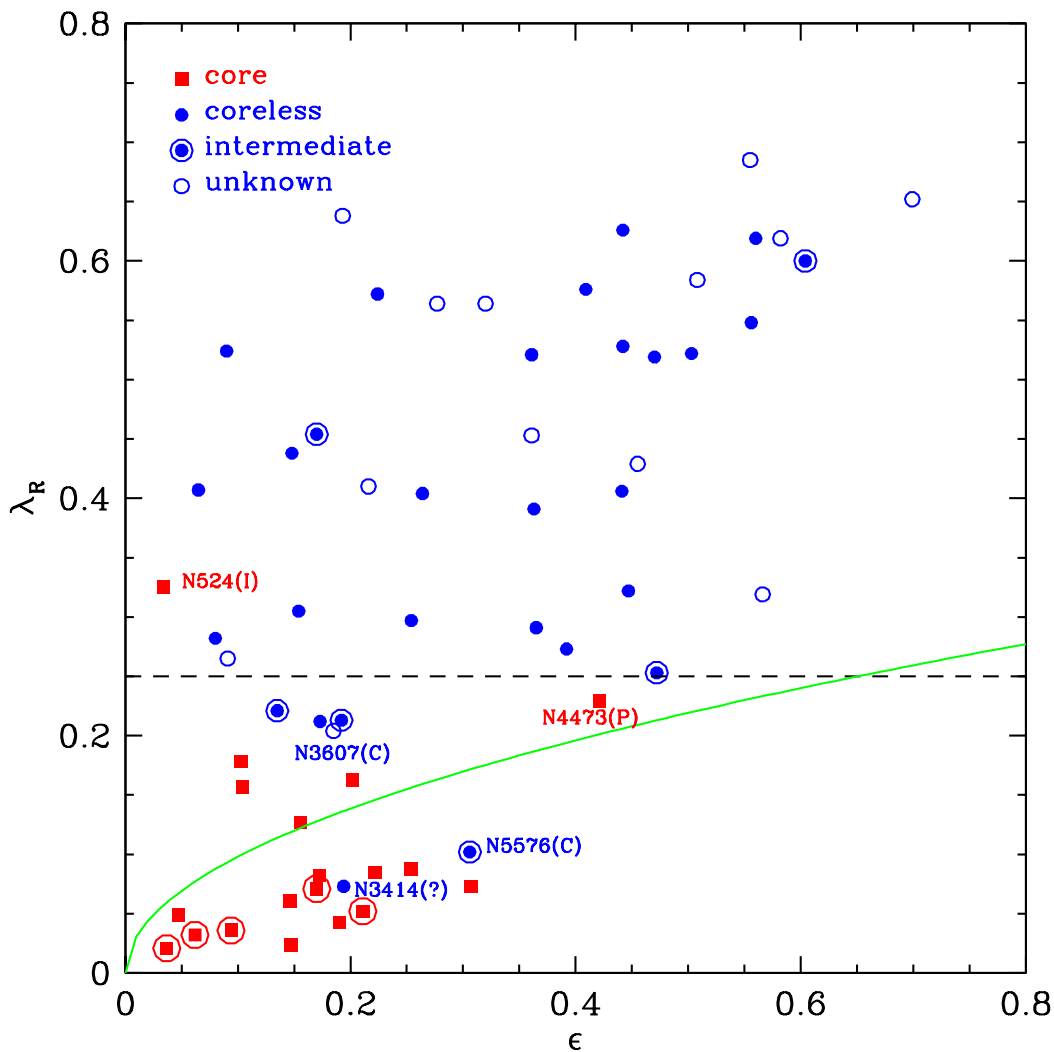

Figure 1. Galaxy rotation parameter $\lambda_R$ is plotted against ellipticity ($\varepsilon$) as measured by Emsellem et al (2011). The core and coreless galaxies (as determined by Krajnovic et al. 2013) are separately marked by red squares and blue circles, respectively. The galaxies with an intermediate profile are marked by large blue open circles. Those with no measurement of the central radial profile structure are marked by small blue open circles. The large red circles indicate 5 cDs. The solid green curve is $\lambda_R = 0.31\ \varepsilon^{0.5}$ used by Emsellem et al. (2011) and Krajnovic et al. (2013) and the dashed line is $\lambda_R = 0.25$ used by Lauer (2012).

The classification of core and coreless galaxies by Kranjnovic et al. (2013) is in most cases consistent with earlier results by Lauer et al. (2007), Hopkins et al. (2009) and Kormendy et al. (2009). However, there are a few notable exceptions, which are marked by galaxy name followed by the different classification in Figure 1. The misclassification, if confirmed, may explain some extreme outliers. NGC 524, the core galaxy with the highest $\lambda_R$ in our sample, was identified as intermediate by Lauer et al. (2007). NGC 4473 was identified as a power-law (coreless) galaxy by Kormendy et al. (2009 – see a rather long discussion on the definition of cores in their section 9.2; see also Dullo & Graham 2013). NGC 5576, a slow rotating galaxy, was classified as intermediate by Kranjnovic et al. (2013), but as a core galaxy by both Lauer et al. (2007) and Hopkins et al. (2009). NGC 3607, with the lowest $\lambda_R$ among those not listed in Kranjnovic et al. (2013), is identified as a core by Lauer et al. (2007) and Hopkins et al. (2009).

If we accept possible misclassifications, the core galaxies in our sample have lower $\lambda_R$ than coreless galaxies, N3607 having the highest $\lambda_R = 0.209$ among core galaxies. The only exception would be NGC 3414, which both Lauer et al. (2007) and Krajnovic et al. (2013) classified as coreless. Given its round shape ($\varepsilon = 0.19$), NGC 3414 may be affected by projection effects, as pointed out by Lauer (2012).

## 3. X-RAY DATA ANALYSIS

X-ray temperatures and luminosities are listed in Table 2. We followed BKF and Kim & Fabbiano (2013), to measure $L_{X,GAS}$ and $T_{GAS}$. We generated light curves to check for background flares and excluded these events (see BKF for more details). The background spectra were then extracted from source-free region within the same CCD. We also use sky background data ((http://cxc.cfa.harvard.edu/ciao/threads/ acisbackground/), after rescaling them by matching the count rates at 9-12 keV (see http://cxc.harvard.edu/contrib/maxim/acisbg/). The two methods give consistent results. In the least X-ray luminous galaxies, the X-ray luminosity of the hot gas may be lower than the integrated contribution of discrete sources, including low-mass X-ray binaries (LMXB) and even fainter X-ray sources, such as active binaries (AB) and cataclysmic variables (CV), see BKF. Therefore, careful determination and subtraction of all stellar contributions is critical to measure $L_{X,GAS}$ accurately. After subtracting the bright LMXBs detected in *Chandra* observations, we fitted the remaining 0.3-5 keV unresolved emission with a 4-component model, consisting of (1) thermal plasma APEC to model the hot gas with metal abundance in the APEC model for the hot gas fixed at solar (Grevesse & Sauval 1998); (2) 7 keV thermal Bremsstrahlung to model the undetected LMXB population; (3) additional APEC component and (4) power-law to model the AB+CV population. In all cases, we find that the spectral determination of the undetected LMXB component is consistent within errors with the estimate derived from extrapolating the X-ray luminosity function of LMXBs (Kim & Fabbiano 2010; see also section 3.4.2 in BKF). To model the AB+CV contribution, we used the spectral parameters determined from the *Chandra* spectra of M31 and M32, where all the LMXBs can be detected and removed (see Appendix in BKF). The normalizations were scaled, based on the K-band luminosity within the region of interest. For extremely gas-poor galaxies where $L_{X,GAS}$ is lower than that of the soft (APEC) component of AB+CV, we considered the error of the soft component of the AB+CV emission and added it in quadrature to the statistical error. The error of the hard component (power-law) of AB+CV is negligible**.** When the gas temperature is not well-constrained, we fixed $T_{GAS}$ at 0.3 keV (a typical value for gas-poor galaxies) to calculate $L_{X,GAS}$. We only used these galaxies in the $L_{X,GAS} - L_K$ correlation study, but not in the $L_{X,GAS} - T_{GAS}$ correlation.

For some well-studied gas rich galaxies in our sample, the emission of the hot gaseous halo extends radially over more than 4 arcminutes, and therefore falls outside the boundaries of a *Chandra* ACIS CCD chip. In these cases (marked in Table 2), we used $L_{X,Total}$ from the *ROSAT* measurements by O'Sullivan et al. (2001) and corrected it for our energy band (0.3-8keV) and the distances in Table 1. We also subtracted the LMXB contribution by applying the BKF scaling relation $L_{X,LMXB}/L_K = 10^{29}$ erg s$^{-1}$ $L_{K\odot}^{-1}$. Because of the scatter in this relation, $L_{X,LMXB}$ may vary by ~50% (BKF). This scatter could cause an error up to 15% in $L_{X,GAS}$. We added this error in the estimates shown in Table 2.

For galaxies with extended hot gas (mostly those for which we use ROSAT data, see Table 2), $T_{GAS}$ may vary as a function of distance from the galaxy center, e.g., $T_{GAS}$ increases from 0.6-0.8 at the central region to 1-1.2 keV at the outskirts (e.g., Diehl and Statler 2008; Nagino & Matshshita 2009). In these cases, we use an emission-weighted average temperature, obtained by fitting the spectra extracted in multiple annuli. For galaxies with extended gas beyond the ACIS chips, we compared our measurements with results in the literature. For example, Nagino & Matsushita (2009) recently analyzed *Chandra* and *XMM-Newton* observations for a sample of gas rich elliptical galaxies and reported their gas luminosities and temperatures in concentric annuli. We estimated the emission-weighted average temperature and confirmed that our results are consistent with those of Nagino & Matsushita within the uncertainties. For the overlapping ETGs, our results are comparable to those in BKF, although should be considered slightly improved because of the improved atomic data in AtomDB (www.atomdb.org) used in the present work, and of the additional data used, when available in the *Chandra* archive.

## 4. THE SCALING RELATIONS

Figure 2 (Left) shows the $L_{X,GAS} - L_K$ scatter plot for our sample of 61 ETGs. As in BKF, we find a good correlation between $L_{X,GAS}$ and $L_K$. The best fit line, $L_{X,GAS} \sim L_K^{3.0\pm0.4}$, is marked by the red dashed line. Figure 2 (Right) shows the $L_{X,GAS} - T_{GAS}$ scatter plot. As in BKF, we find a positive correlation between $L_{X,GAS}$ and $T_{GAS}$, with best fit relation is $L_{X,GAS} \sim T_{GAS}^{5.4\pm0.5}$. The distribution of the five cD-type galaxies in this plot is consistent with the $L_{X,GAS} - T_{GAS}$ of group/cluster dominant galaxies (O'Sullivan et al. 2003; the green line) which is shifted upward toward higher luminosities for a given gas temperature relative to non-cD E galaxies (see section 5.2 for more discussions).

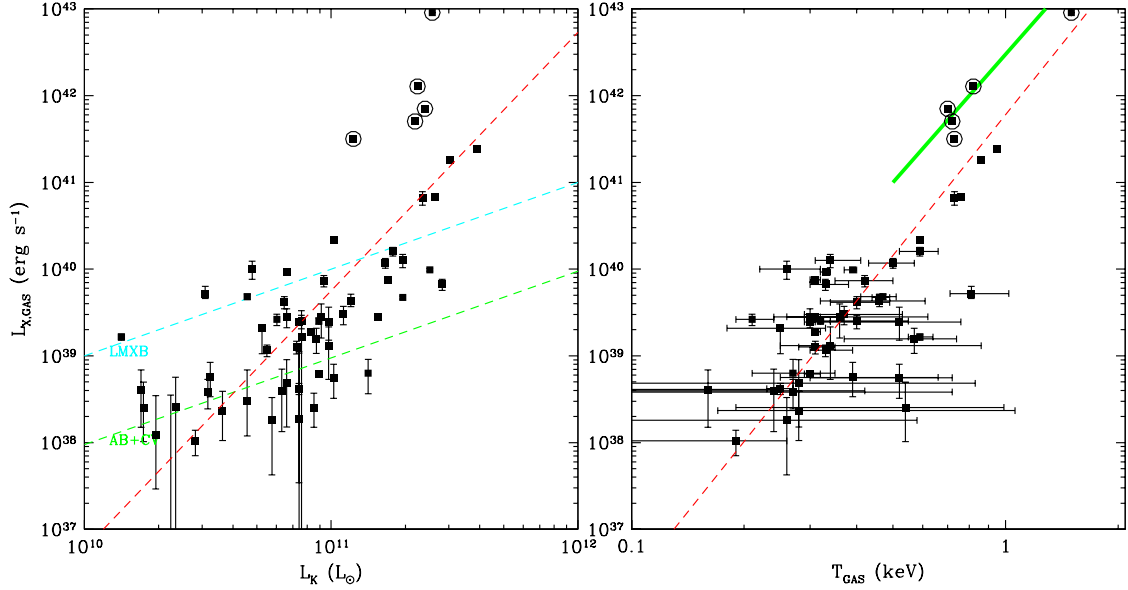

Figure 2. **Left -** $L_{X,GAS}$ plotted against $L_K$ for early type galaxies selected from the ATLAS$^{3D}$ sample. The cyan and green lines in indicate the BKF linear relations for $L_{X,LMXB}$ and $L_{X,AB+CV}$, respectively, and the red line ($L_{X,GAS} \sim L_K^3$) is the best fit relation for the entire sample. The five galaxies at the highest $L_X$ marked by a big circle are cD galaxies. **Right -** $L_{X,GAS}$ plotted against $T_{GAS}$ for early type galaxies selected from the ATLAS$^{3D}$ sample, with cD galaxies marked by a big circle. The red dashed line is the best fit for the entire sample ($L_{X,GAS} \sim T_{GAS}^{5.4}$). The thick green line is the relation determined among group/cluster dominant galaxies by O'Sullivan et al. (2003

Figures 3 and 4 show the $L_{X,GAS} - L_K$ and $L_{X,GAS} - T_{GAS}$ scatter plots, for different classes of ETGs, based on the properties of the ATLAS$^{3D}$ sample listed in Table 1: (a) morphological types (E vs. S0), (b) central stellar velocity dispersion (high vs. low $\sigma$), (c) central radial profile (core vs. coreless), (d) mean stellar age (old vs. young), (e) flattening (round vs. flat), and (f) stellar kinematics (slow vs. fast rotators). As noticed by Kormendy et al. (2009; see also Lauer 2012), core galaxies (red points in Figure 3a-f), are characterized by large $L_K$, high stellar velocity dispersion ($\sigma$), round isophotes, no sign of recent star formation (SF) and slowly rotating stellar kinematics. These galaxies also tend to have larger $L_{X,GAS}$ (see also, Fabbiano 1989; Bender et al. 1989; Eskridge et al. 1995). Coreless galaxies (blue points) are characterized by the opposite properties. While some core (red) galaxies can be hot gas poor, with $L_{X,GAS}$ as low as a few $10^{38}$ erg s$^{-1}$, coreless galaxies are all hot gas poor (as previously reported, e.g., Pellegrini 1999). Figure 4 shows that the $L_{X,GAS}$ - $T_{GAS}$ relation is very tight for core galaxies (red points), but there is no correlation for coreless galaxies.

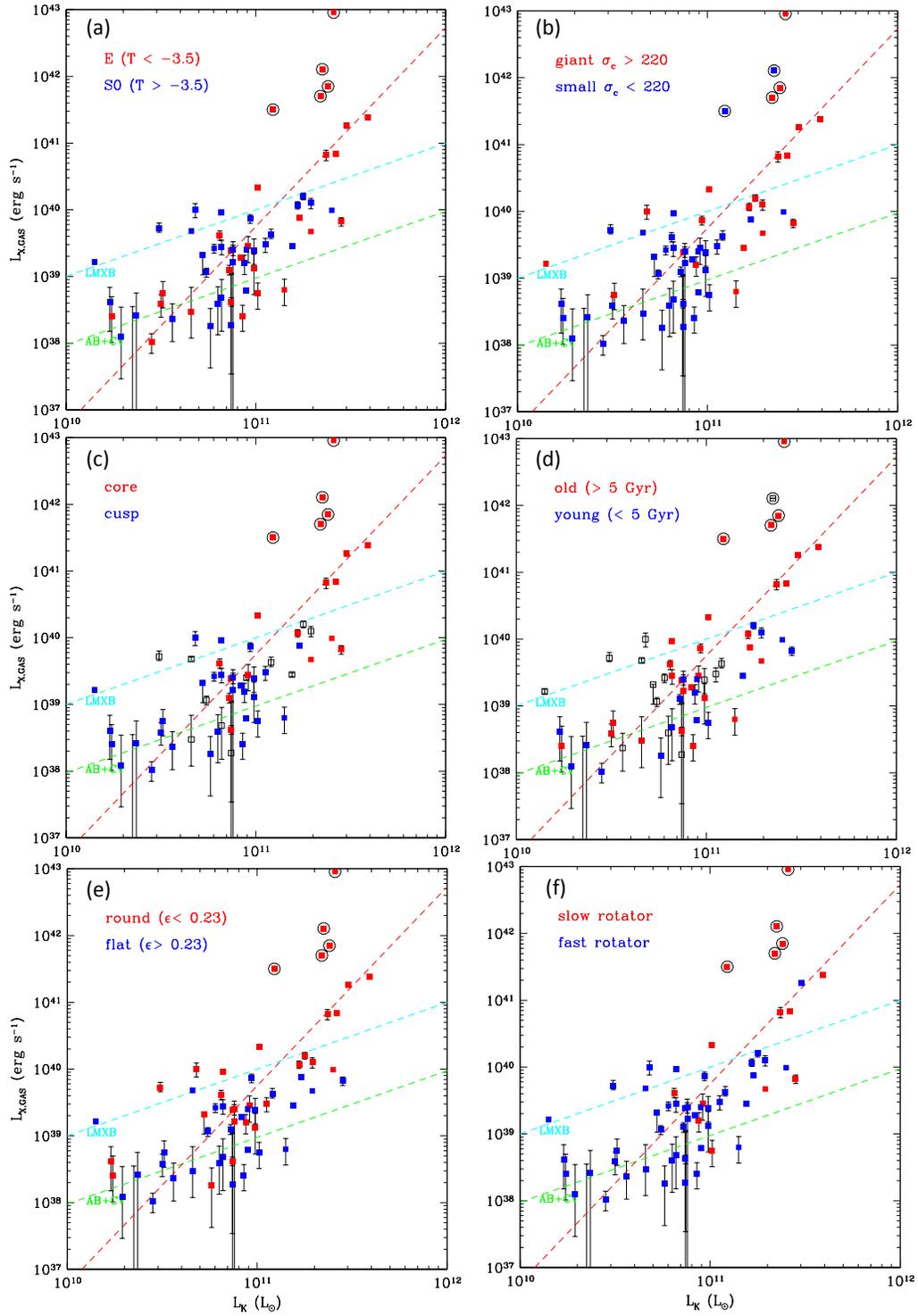

Figure 3. Same as Figure 1a, but with two classes of ETGs marked with different colors, based on (a) their morphological types, (b) central stellar velocity dispersion, (c) central radial profile, (d) mean stellar age, (e) flattening, and (f) stellar rotation.

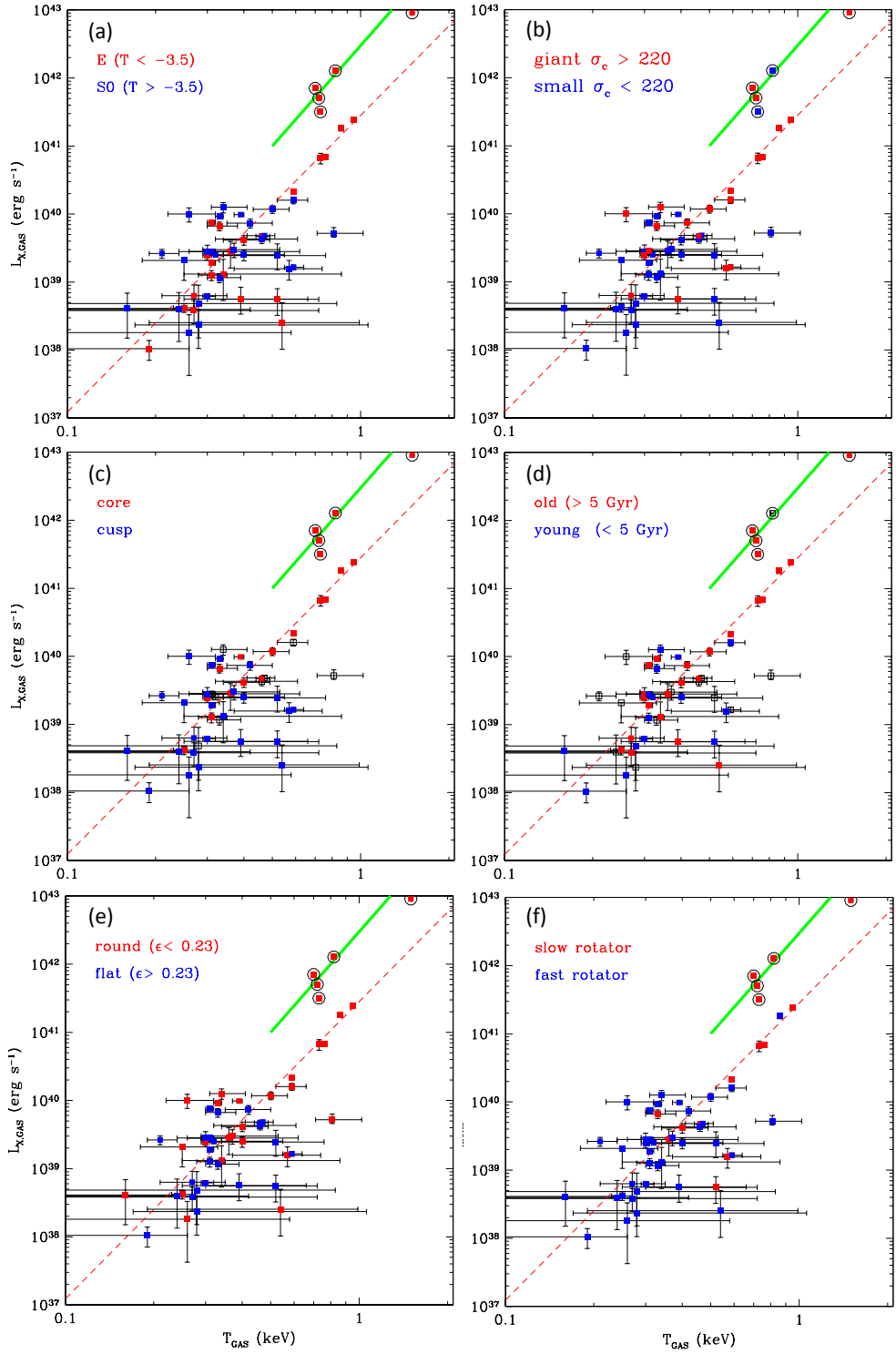

Figure 4. Same as Figure 2b, but with two classes of ETGs marked with different colors (same as Figure 3). The red dashed line is the best fit relation among the core galaxies ($L_{X,GAS} \sim T_{GAS}^{4.4}$).

To determine the best-fit relations, we have applied a bisector linear regression method and estimated the corresponding error by bootstrap resampling (http://home.strw.leidenuniv.nl/~sifon/pycorner/bces). We also used the Pearson and Spearman correlation tests from the *scipy* statistics package (http://www.scipy.org) to estimate the p-value for the null hypothesis. The results are summarized in Tables 3 and 4 for the $L_{X,GAS}$ - $L_K$ and $L_{X,GAS}$ – $T_{GAS}$, respectively. In both Tables, we list the results for the full sample, and subsamples with different central structure.

4.1 $L_{X,GAS}$ - $L_K$

For the entire sample, $L_{X,GAS} \sim L_K^{3.0\pm0.4}$. The best fit relation becomes steeper for core galaxies, when cDs are included, with $L_{X,GAS} \sim L_K^{4.3\pm0.7}$; it is otherwise consistent with that of the full sample. The relation is significantly flatter for coreless galaxies with $L_{X,GAS} \sim L_K^{1.6\pm0.3}$. The p-values determined by Pearson and Spearman tests for the full sample are very low (~$10^{-10}$), indicating a strong correlation. The correlations are less strong, but still significant, in the sub-samples, particularly in the coreless sample (p-value ~ 0.01).

The scatter of the correlation is large in all cases, as indicated by the rms deviations reported in Table 3: 0.8 dex in $L_{X,GAS}$ for a given $L_K$, for the entire sample and 0.5 dex for the core (without cDs) and coreless sub-samples. For example, at $L_K = 10^{11}$ $L_{K_\odot}$, which is close to the characteristic luminosity ($L*$) of the galaxy luminosity function, (e.g., Cole et al. 2001), NGC 4621 and NGC 4636 have similar K-band luminosities (1.4 and 1.2 x $10^{11}$ $L_{K_\odot}$, respectively), but their $L_{X,GAS}$ differ by a factor of 500. At the higher $L_K$ end, NGC 5322 and NGC 4486 have similar $L_K$ (~3 x $10^{11}$ $L_{K_\odot}$) but their $L_{X,GAS}$ differ by a factor of 1400. Note that NGC 4636 and NGC 4486 are cD galaxies. Excluding these cDs, the spread in $L_{X,GAS}$ is somewhat reduced to a factor of ~30 in $L_{X,GAS}$ for a given $L_K$ in the entire $L_K$ range seen in Figure 2a.

4.2 $L_{X,GAS}$ – $T_{GAS}$

For the entire sample, $L_{X,GAS}$ and $T_{GAS}$ are well correlated with p-value < $10^{-8}$ (Table 4). The best fit relation is $L_{X,GAS} \sim T_{GAS}^{5.4\pm0.6}$, slightly steeper than that of BKF ($L_{X,GAS} \sim T_{GAS}^{4.5}$). This is mainly because of the cD galaxies included in our sample. The relation is particularly tight ($L_{X,GAS} \sim T_{GAS}^{4.4\pm0.3}$) for the normal (non-cD) core galaxies, with a small normalization error and only 0.2 dex rms. This tight relation holds for a range of $kT_{GAS} = 0.3 - 1$ keV and $L_{X,GAS} =$ a few x $10^{38}$ – several x $10^{41}$ erg s$^{-1}$.

We further checked the core galaxy sample for signatures of recent star formation. NGC 524 is classified as S0 by Lauer (2007) and Kormendy and Ho (2013) as well as by RSA and RC3. Furthermore molecular gas (M(H$_2$) = $10^8$ M$_\odot$) was detected in this galaxy by the Atlas$^{3D}$ team (see Table 1). NGC 4382 is classified as S0 by Lauer et al. (2007), RSA and RC3, but as E by Hopkins et al (2007) and Kormendy and Ho (2013). No atomic and molecular gas was detected in this galaxy. However, its average stellar age is quite young, 1.6 Gyr (see Table 1). NGC 5322 is unanimously classified as E, but its stellar age is 2.4 Gyr (see Table 1). Excluding these three galaxies, all the remaining core galaxies are classified as E, their stellar ages are old and no cold gas and dust are detected. With NGC 524 excluded, this sample consists of only slowly rotating galaxies. We can consider them as genuine, passively evolving (no recent star formation), σ-supported (slow or no rotation), morphologically elliptical galaxies. Table 4 shows an even tighter $L_{X,GAS}$ – $T_{GAS}$ correlation for this genuine elliptical sample, with rms deviation of only 0.13 dex.

In contrast, no correlation exists for coreless galaxies (p-value ~ 1). Given that coreless ETGs are generally hot gas poor, their errors (particularly in $T_{GAS}$) are large. To quantitatively test the possibility that a tight correlation may be hidden because of the large errors, we ran simulations by redistributing points in the $L_{X,GAS}$-$T_{GAS}$ plane, assuming a Gaussian distribution with σ equal to the error of each point. In 100 simulations, we found one (or zero) incidence where the p-value is less than the conventionally used limit of 0.05 (or 0.01), indicating that it is unlikely that the observational errors could hide a real correlation.

Table 3
Best fit $L_{X,GAS}$ - $L_K$ Relation

| sample | N | A | err | B | err | p-values Pearson | Spearman | rms deviation in log($L_{X,GAS}$) |
|---|---|---|---|---|---|---|---|---|
| (1) full | 59 | 2.98 | 0.36 | -0.25 | 0.11 | 3x10-10 | 9x10-11 | 0.79 |
| (2) core | 19 | 4.27 | 0.72 | -0.41 | 0.21 | 0.002 | 0.004 | 0.91 |
| (3) core-cDs | 14 | 2.84 | 0.43 | -0.52 | 0.17 | 0.0007 | 0.0004 | 0.50 |
| (4) noncore | 40 | 1.83 | 0.22 | -0.54 | 0.09 | 0.0006 | 0.0008 | 0.55 |
| (5) coreless | 23 | 1.60 | 0.28 | -0.64 | 0.12 | 0.008 | 0.01 | 0.47 |

$\log(L_{X,GAS} / 10^{40}$ erg s$^{-1}) = A \log(L_K / 10^{11} L_{K_\odot}) + B$

(1) the entire sample
(2) core galaxies only  (1 in Table 1 column 9)
(3) same as (2) but also exclude 5 cD galaxies
(4) all but core galaxies (2, 3, and 9 in Table 1 column 9)
(5) only coreless galaxies (3 in Table 1 column 9)

Table 4
Best fit $L_{X,GAS}$ − $T_{GAS}$ Relation

| sample | N | A | err | B | err | p-values Pearson | Spearman | rms deviation in log($L_{X,GAS}$) |
|---|---|---|---|---|---|---|---|---|
| (1) full | 49 | 5.39 | 0.60 | 0.16 | 0.43 | 3x10-11 | 1x10-8 | 0.72 |
| (2) core | 19 | 6.16 | 0.70 | 0.34 | 0.11 | 2x10-8 | 4x10-7 | 0.47 |
| (3) core-cD | 14 | 4.36 | 0.30 | 0.14 | 0.06 | 2x10-8 | 3x10-7 | 0.20 |
| (4) core-cD-SF | 11 | 4.57 | 0.26 | 0.07 | 0.05 | 1x10-8 | 4x10-9 | 0.13 |
| (5) noncore | 30 | -- | -- | -- | -- | 0.37 | 0.34 | -- |
| (6) coreless | 16 | -- | -- | -- | -- | 0.94 | 0.99 | -- |

$\log(L_{X,GAS} / 10^{40}$ erg s$^{-1}) = A \log(kT_{GAS} / 0.5$ keV$) + B$

(1) the entire sample
(2) core galaxies only  (1 in Table 1 column 9)
(3) same as (2) but also exclude 5 cD galaxies
(4) same as (3) but also exclude 3 possibly star forming galaxies
(5) all but core galaxies (2, 3, and 9 in Table 1 column 9)
(6) only coreless galaxies (3 in Table 1 column 9)

# 5. DISCUSSION

Using the well-studied ATLAS$^{3D}$ sample of 61 ETGs, we have rebuilt the scaling relations, $L_{X,GAS}$ – $L_K$ and $L_{X,GAS}$ – $T_{GAS}$, and investigated their behavior in the subsamples of core and coreless galaxies. For core E galaxies, we find that the $L_{X,GAS}$ – $T_{GAS}$ relation is significantly tighter than the $L_{X,GAS}$ – $L_K$ one. In particular, when we eliminate galaxies with possible relatively recent star formation, yielding a sample of genuine old core ellipticals, we find the tightest $L_{X,GAS}$ – $T_{GAS}$ correlation. For coreless galaxies, instead, we find a weak $L_{X,GAS}$ – $L_K$ correlation, while $L_{X,GAS}$ – $T_{GAS}$ are not correlated. Below we discuss the implications of these results.

## 5.1 The $L_{X,GAS}$ – $T_{Gas}$ correlation of Genuine Core Elliptical Galaxies

As shown in Section 4.2, genuine, passively evolving (no recent star formation), σ-supported (slow or no rotation), core, non-cD, morphologically elliptical galaxies present the strongest correlation between gas luminosity and temperature. This is tightest among all the X-ray scaling relations ever reported for ETGs (e.g., Fabbiano 1989; Bender et al. 1989; Eskridge et al. 1995; Mathews & Brighenti 2003; BKF). Both best-fit slope and normalization are well determined, and the rms deviation is only 0.13 dex:

$$\log(L_{X,GAS} / 10^{40} \text{ erg s}^{-1}) = (4.6 \pm 0.3) \times \log(kT_{GAS} / 0.5 \text{ keV}) + (0.07 \pm 0.05).$$

The tight correlation indicates that the ability to retain hot gas ($L_{X,GAS}$), which is a function of the potential depth, and the balance ($T_{Gas}$) between heating and cooling are closely regulated. We can qualitatively understand this correlation as a consequence of virialized gaseous halos in the dark matter potentials of these galaxies (e.g., Mathew et al. 2006; Kim & Fabbiano 2013). Larger galaxies not only retain larger amounts of hot ISM, but also add more energy to the ISM from stellar and AGN feedback (e.g., Mathew & Brighenti 2003; Pellegrini 2011; Pellegrini et al 2012).

This relation is tight and steep. At the low end, we have NGC 3399 with $kT_{GAS} = 0.3$ keV and $L_{X,GAS} = 4 \times 10^{38}$ erg s$^{-1}$, while at the high end, NGC 4472 and NGC 4649 with $L_{X,GAS} = 2 \times 10^{41}$ erg s$^{-1}$ and $kT_{GAS} = 0.9$ keV. In Figure 5 (Left), we show the $L_{X,GAS}$ - $T_{GAS}$ scatter plot for the genuine E sample. In Figure 5 (Right) we show the $L_{X,GAS}$ – $T_{GAS}$ relation recently determined by Negri et al. (2014) who performed high resolution 2D hydrodynamical simulations separately for fully σ-supported and rotation-supported galaxies. The similarity between the observation of genuine Es (Figure 5 Left) and the simulation of σ-supported Es (red in Figure 5 Right) is remarkable for hot-gas rich Es with $L_{X,GAS} > 10^{40}$ erg s$^{-1}$. In particular, the slope is well matched. The normalization is offset in that $L_{X,GAS}$ is slightly higher for a given $T_{GAS}$ in the simulations than in the observations, but still within twice the rms deviation. The difference may be caused by the uncertainties in various parameters involved and other effects not considered in the simulations (e.g., AGN feedback, magnetic pressure and turbulence etc.).

If AGN feedback is a sporadic event, our result (the tight relation in a wide range of $L_{X,GAS}$) may suggest that the AGN feedback is not critical among the genuine Es, also shown in the above simulation without AGN feedback. Alternatively, AGN feedback could work continuously in a radio mode by smoothing episodic energy input inside the hot gas rich normal Es (Kormendy et al. 2009) and may also scale with the halo mass as suggested by Booth & Schaye (2010) and Bogdan & Goulding (2015).

What is surprising, in comparison with the simulation results, is that the observed tight $L_{X,GAS}$-$T_{GAS}$ relation holds for the entire sample of genuine Es, extending down to X-ray luminosities in the range of a few $10^{38}$ erg s$^{-1}$. The simulations, instead, predict higher gas temperatures at low $L_X$, both for σ-supported and rotation-supported galaxies (see Figure 5b, where σ-supported Es are not well represented, but the range of gas temperatures is a robust prediction of Negri et al.). In fact, the ISM of gas-poor Es is likely to be in the outflow/wind state, and then tends to be at a temperature higher than extrapolated from the $L_{X,GAS}$ - $T_{GAS}$ relation valid for inflows (the typical flow phase of gas-rich Es; e.g., Pellegrini 2011, Negri et al. 2014). The presence of a tight correlation extending to these lower X-ray luminosities suggest that the hot ISM may be retained in equilibrium in these galaxies and may require a lower rate of the energy input (e.g., lower SNe rates, reduced AGN cycles).

We note that we cannot directly compare the simulations of isotropic rotators with the observations of coreless galaxies (or more specifically fast rotators), because the physical state of their hot ISM may be affected by multiple factors (see section 5.3). However, the upper reaches of $L_{X,GAS}$ for this subsample are consistent with the distribution of the isotropic rotator models (see Figure 4).

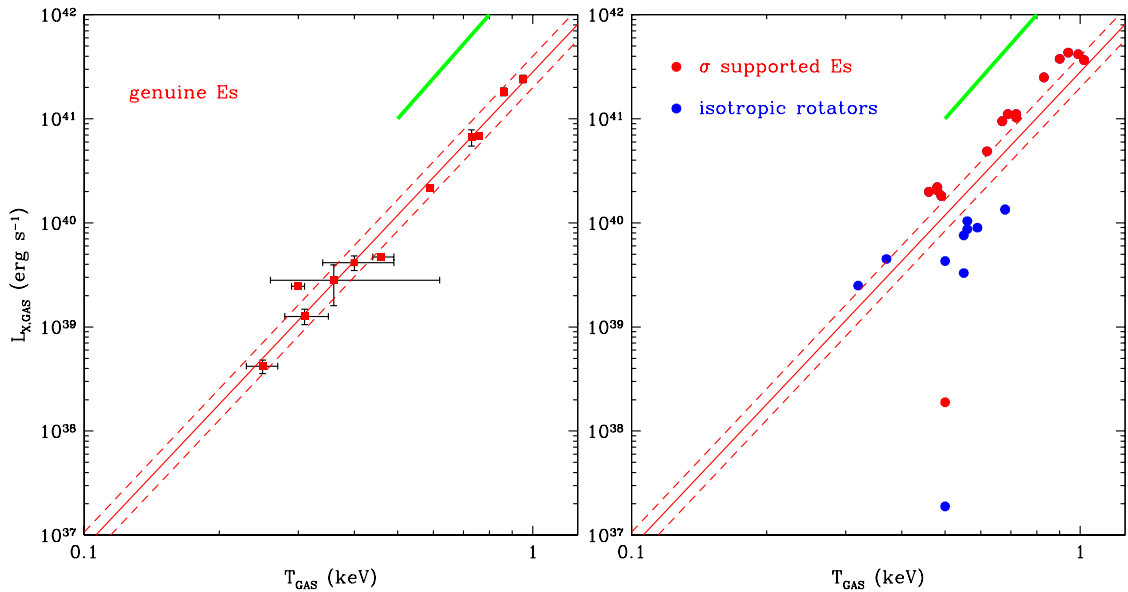

Figure 5. $L_{X,GAS}$ vs. $T_{GAS}$ for the genuine E sample (Left) from observations and (Right) from simulations (Negri et al. 2014). The diagonal lines are the best fits (solid line) with rms deviations (dashed lines).

**5.2 Giant Ellipticals versus cD Galaxies**

In our sample, we have five cD-type galaxies, NGC 4406, NGC 4486, NGC 4636, NGC 5813, and NGC 5846. We have found that they have an order of magnitude higher $L_{X,GAS}$ for a given $L_K$ or $T_{GAS}$ (Figure 2), compared with normal giant E galaxies, and follow the same $L_{X,GAS} - T_{GAS}$ relation of similar group/cluster dominant galaxies by O'Sullivan et al. (2003). This difference may be related to the associations of cDs with the larger extended group or cluster dark matter halos. NGC 4486 (M87) is in the center of the Virgo cluster. NGC 4636, NGC 5846 and NGC 5813 are dominant galaxies in groups with 10-20 known members in Nearby Galaxies Catalog (Tully 1988) and in the 2MASS group catalog (Crook et al. 2008). NGC 4406 (M86) is not identified as a group dominant galaxy in both catalogs, However, it is the biggest among several nearby galaxies at the similar 3D distance, including NGC 4374, NGC 4388, NGC 4438, and NGC 4458 and also at the center of a clump of dEs with negative velocities (M86 velocity = -250 km s$^{-1}$), forming a second massive sub-clumps outside a similar clump centered on M87 (e.g., Binggeli et al 1993). Therefore, it is likely that M86 is sitting at the bottom of a group potential well. Alternatively, if the X-ray emission from the extended plum on the NW side of the galaxy center (e.g., Randall et al. 2008) is excluded, the X-ray luminosity within the main galaxy body (r < 3') decreases by a factor of 4, while $T_{GAS}$ (0.8 keV) remains similar (e.g., Rangarajan et al. 1995). Then its $L_{X,GAS}$ and $T_{GAS}$ would be close to those of normal giant Es.

The gas luminosity (or gas mass), being larger in the central dominant galaxies in groups and clusters, does reflect the total mass. This is consistent with the tight correlation between the total mass and $L_{X,GAS}$ found by Kim & Fabbiano (2013). Qualitatively, the larger amount of dark matter in cDs may work in two ways to increase $L_{X,GAS}$: (1) by increasing the ability to hold the hot ISM in the deeper potential well, and (2) by adding external pressure from ICM/IGM to confine the hot ISM. The former would also increase the gas temperature (e.g., Negri et al. 2014) if cooling is not enhanced (see below), while the latter could increase $L_{X,GAS}$ without significantly affecting $T_{GAS}$.

To better understand the difference between cDs and giant Es, we increase the sample of high $L_X$ galaxies with five additional well-studied hot-gas-rich elliptical galaxies not included in the ATLAS$^{3D}$ sample (see Figure 6). Their properties are summarized in Table 5. The X-ray properties are extracted from the literature (see references in the table). $T_{GAS}$ is estimated as a luminosity-weighted average temperature in the same way as in our data analysis described in section 3. Three (NGC 507, NGC 5044, NGC 7619) of these galaxies are cD-type galaxies in groups with 10-30 members (Crook et al. 2007) and are consistent with other cDs of the ATLAS$^{3D}$ sample, but two (NGC 1399 and NGC 1407) follow the relation of giant core E galaxies; of these only NGC 1399 has been classified as a cD (e.g., Schombert 1986) in the center of the Fornax cluster.

Werner et al. (2014) recently reported that the entropy profiles of the hot halos of giant E and cD galaxies show a dichotomy. The galaxies displaying extended optical nebulae as seen in [CII] emission lines have lower entropies beyond r > 1 kpc than cold gas poor galaxies. It is interesting to note that galaxies with lower entropies are mostly cDs while those with higher entropies are normal core Es in our scaling relations. Three (NGC 4636, NGC 5813, NGC 5846) of the five galaxies with extended [CII] emission in Werner et al. (2014) are in our sample and they are all cDs with excess $L_{X,GAS}$ for their $L_K$ and $T_{GAS}$. Of the remaining two galaxies, NGC 5044 follows the same trend as the other cDs (see Figure 6, Right) while NGC 6868 does not (it was recognized as a different case with a rotating disc by Werner et al. 2014, see their section 4). On the other hand, three (NGC 4472, NGC 4649, NGC 4261) of the five galaxies with no [CII] emission in Werner et al. (2014) are in our sample and they are normal core Es. We also plot the renaming two galaxies in our scaling relation. NGC 1407, a core giant E, follows the same trend of normal core E as expected. However, NGC 1399, a well-known central dominant galaxy (cD) in the Fornax cluster, also behaves like a normal core E (see below).

Because the gas temperatures vary only within a small range (0.8 - 1 keV), the entropy ($k_BT/n_e^{2/3}$) dichotomy (at r > 1 kpc) must be caused by the gas density ($n_e$), i.e., cDs with lower entropies having higher densities (hence higher $L_{X,GAS}$) than normal core E galaxies, as expected with an additional confinement by ICM/IGM. Werner et al. (2014) further suggested that cooling from the hot phase mainly produced cold gas in galaxies with extended [CII] emission. This can be understood because cDs have a higher density where the cooling rate increases with $n_e^2$ and also the thermal stability parameter ($\sim T/n_e^2$) decreases (the gas becomes thermally unstable – see Figure 10 in Werner et al. 2014). The enhanced cooling in higher density (lower entropy) gas in cDs may explain why their gas temperature is almost the same as that of giant Es, while their luminosity is higher by an order of magnitude. A quantitative study is warranted to better understand these observed differences between giant Es and cDs.

NGC 1399 is the one exception. While its location in the $L_{X,GAS}$ - $T_{GAS}$ plane is consistent with having no extended cold gas (as other core Es), it is a well-known cD. It may be related to an unusual distribution of DM, in the scale comparable to the hot gas extent. The sample of Nagino & Matsushita (2009) includes six hot gas rich galaxies, shown in our Figure 6: three (NGC 4636, NGC 5044, NGC 5846) with extended cold gas and three (NGC 1399, NGC 4472, NGC 4649) without it. While the mass-to-light ratio $(M/L_K)$ is similar in the inner region (< 0.5 $r_e$), this ratio increases more rapidly with increasing radius in those with extended cold gas where $M/L_K$(< 6 $r_e$) is higher by a factor of 8-16 than $M/L_K$(< 0.5 $r_e$), while it only increases by a factor of 4-6 in the other three galaxies (including NGC 1399), indicating that the amount and distribution of DM is important to determine the amount of hot gas properties and its cooling.

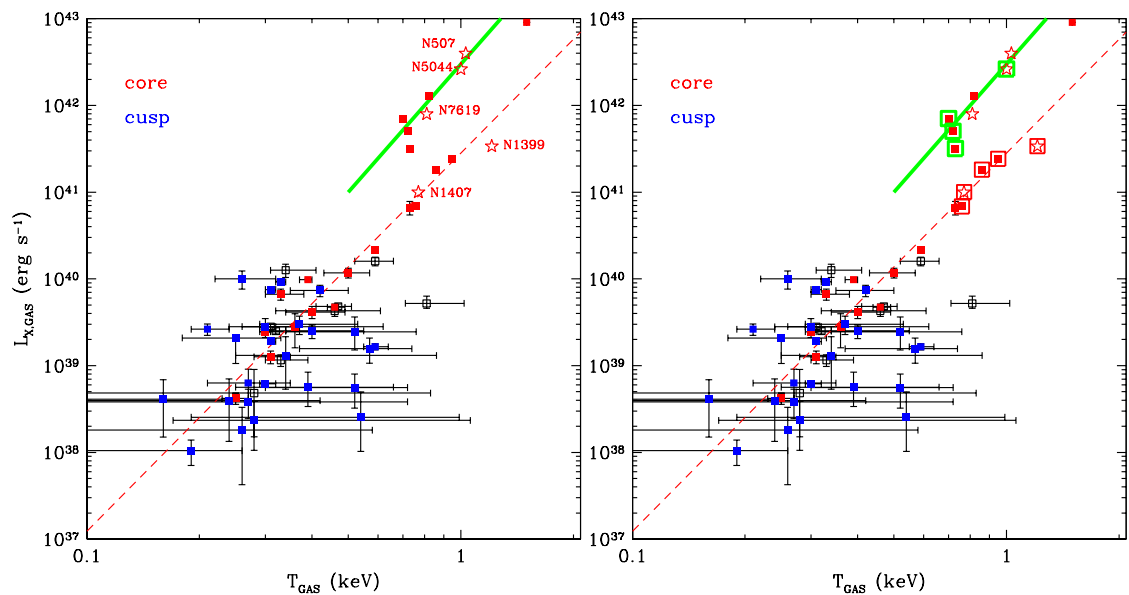

Figure 6. Left - Same as Figure 3c but with 5 more well-studied galaxies (stars with galaxy names) taken from the literature (see Table 5). Right – High luminosity **cD** galaxies with extended [CII] emission are marked by big green squares, while galaxies **(mostly giant Es)** lacking this emission are identified by large red squares.

```
 Table 5 X-ray Properties of Other Galaxies
 ------------------------------------------------------
 name    d(Mpc)   log(L_K)   log(L_X,GAS)   T_GAS   ref
 ------------------------------------------------------
 N0507   69.90    11.69      42.59          1.03    O03
 N1399   19.95    11.40      41.53          1.21    NM09
 N1407   28.84    11.57      41.00          0.77    F06
 N5044   31.19    11.23      42.42          1.00    NM09
 N7619   52.97    11.56      41.89          0.81    O03
 ------------------------------------------------------
References.
O03 - O'Sullivan, E., Ponman, T. J., & Collins, R. S. 2003, MNRAS, 340, 1375
F06 - Fukazawa, Y., et al. 2006, ApJ 636, 698
NM09 - Nagino, R. & Matsushita, K. 2009, A&A 501, 157
```

Giant ellipticals and central dominant cDs particularly in low-mass groups are not always clearly distinguished. The upward shift in the X-ray scaling relations (higher $L_{X,GAS}$ for a given $L_K$ and $T_{GAS}$) can provide an excellent empirical method to correctly identify them, i.e., the presence or absence of the extended DM associated with IGM/ICM.

## 5.3 X-ray Scaling Relations of Coreless ETGs

As discussed above, the X-ray scaling relations for coreless galaxies present a large scatter. $L_{X,GAS}$ and $L_K$ are weakly correlated among coreless galaxies, indicating that the small amount of hot gas near the central region is related to the potential depth determined mostly by stars and to the gas production rate again determined by the stellar mass loss. However, $L_{X,GAS}$ and $T_{GAS}$ are not correlated at all. That is in contrast with core galaxies, which have a strong correlation.

The lack of a $L_{X,GAS}$ - $T_{GAS}$ correlation among coreless galaxies may be understood due to the presence of several factors that may affect the retention and temperature of the hot gas. These galaxies are fast rotators with flattened galaxy figures. Both effects may impact the accumulation of hot halos, as shown by the simulations of Negri et al. (2014). They also contain embedded disks with recent star formation, which may increase the effect of the stellar feedback, increasing the gas temperature (Negri et al. 2015). Each effect can change the hot gas properties by altering the potential field and energy budget. A similar situation may happen in the ISM of spiral galaxies. In Figure 7 we compare the $L_{X,GAS}$ and $T_{GAS}$ relation among the sample of spiral galaxies taken from Li and Wang (2013). Remarkably the parameter space occupied by the spiral galaxies is similar to that of coreless ETGs. In both cases, there is no clear positive correlation.

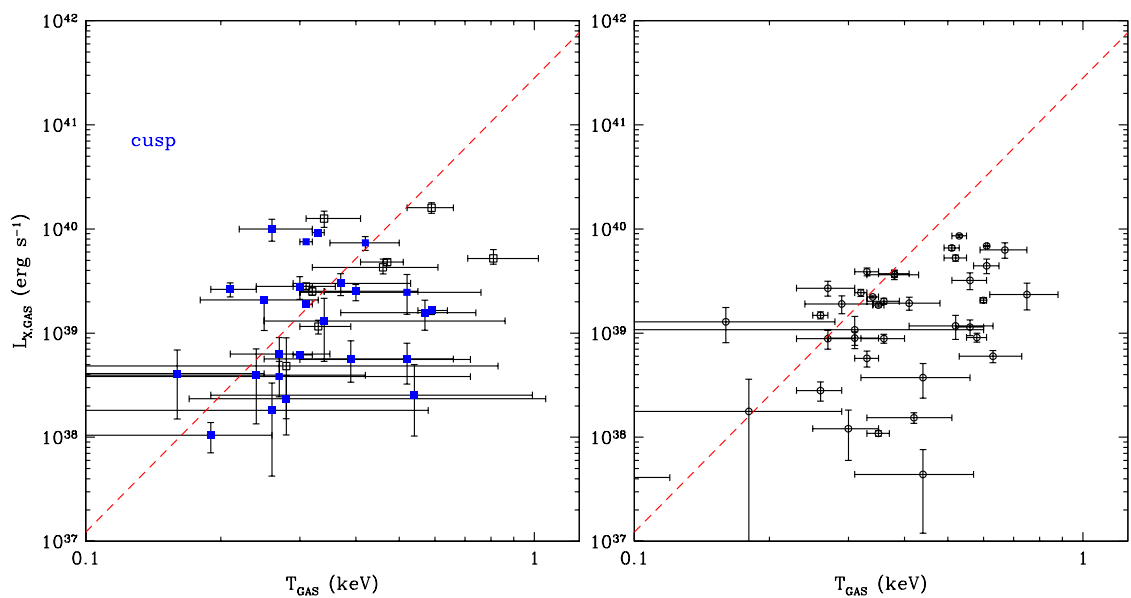

Figure 7. $L_{X,GAS}$ vs. $T_{GAS}$ (a) for a ETG sample without core ellipticals (same as Fig 3c without red points) (b) for a spiral sample (from Li and Wang 2012).

## 5.4 Comparison with Clusters and Groups

In Figure 8, we introduce a schematic diagram to compare the $L_{X,GAS} - T_{GAS}$ relations in the ETG samples we have studied with those reported for cDs (O'Sullivan et al. 2003), groups (Ponman et al. 2003), and cluster of galaxies (Pratt et al 2009). To first approximation, the bigger the system, the hotter and more luminous the gas. The details are more complex, though. Starting from the bottom left corner, the coreless ETGs and spiral galaxies show no clear correlation (see Figures 4c and 7). On the other hand, genuine Es show a tight correlation with a slope of 4.5 ± 0.3 (see Figures 4c and 5). Also this tight relation is well reproduced among σ-supported, hot gas rich elliptical galaxies by high-resolution simulations (Negri et al. 2014). The cDs follow a similar trend as the core Es, but are shifted toward higher $L_{X,GAS}$ values for a given $T_{GAS}$. The larger dark matter halo in which these cDs are embedded may be responsible for the retention of large hot gaseous halos (larger $L_X$). However, the temperature of this gas may not increase because of the presence of cooling connected with the its larger densities (see section 5.3). The groups also have a similar trend, but they are further shifted upward. The clusters at the top right corner have a strong, but flatter relation ($L_{X,GAS} \sim T_{GAS}^3$). As the exact difference between groups and clusters is somewhat subtle, the distinction in the $L_{X,GAS} - T_{GAS}$ relation of groups and clusters may be ambiguous among the big groups and small clusters, it is possible that the slope continuously changes from 4.5 to 3 (see Sun et al. 2009).

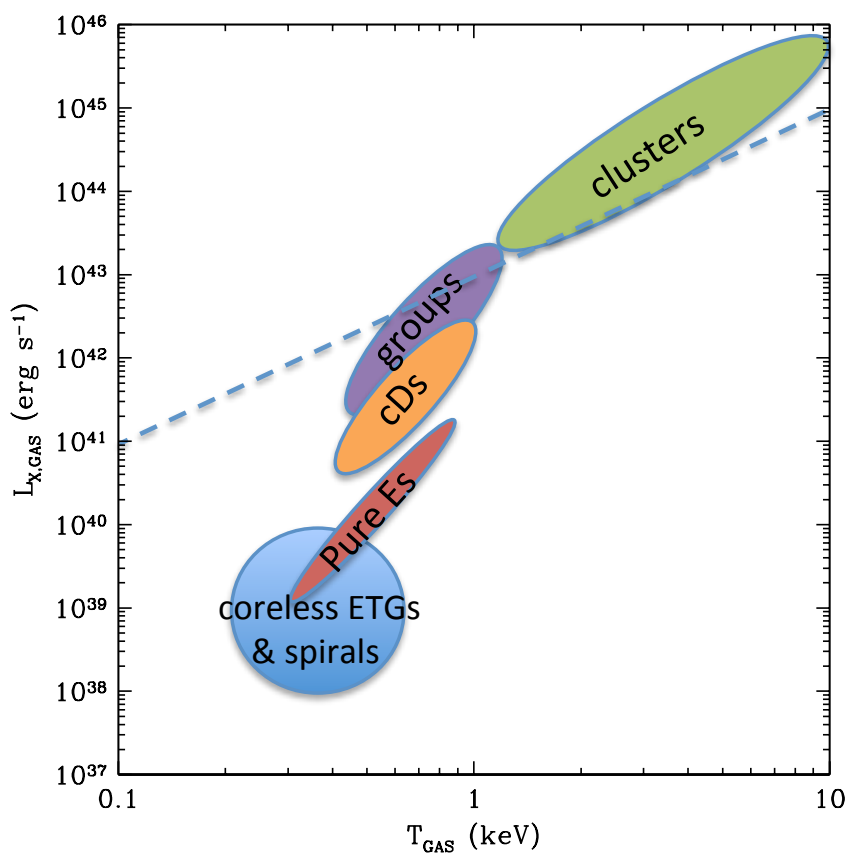

Figure 8. Comparison of the $L_{X,GAS}$ - $T_{GAS}$ relations in various samples. From the bottom left, the coreless ETGs and spirals have no correlation, while the normal (non-CD) core E galaxies have a very tight correlation ($L_{X,GAS} \sim T_{GAS}^{4.5}$). The cDs and groups have a similar trend with the core Es, but they are shifted toward higher $L_{X,GAS}$. The clusters at the top right corner have a flatter relation ($L_{X,GAS} \sim T_{GAS}^3$) that other sub-samples. For a reference, the self similar expectation ($L_{X,GAS} \sim T_{GAS}^2$) is shown in dashed lines.

The $L_{X,GAS}$ - $T_{GAS}$ relation expected by the self similar case (where gravity dominates) has a slope of 2 (dashed lines in Figure 8). The steep slope (3) in clusters indicates that baryonic physics is already important even in this largest scale. In galaxies, the slope is even steeper (4.5), further indicating the increase of importance of non-gravitational effects (including star formation, AGN and their feedback).

## 6. SUMMARY

(1) Using the well-characterized ATLAS$^{3D}$ sample of 61 ETGs, we confirm our previous results (BKF) of the existence of two correlations in ETG samples: $L_{X,GAS} - L_K$ and $L_{X,GAS} - T_{GAS}$. The best fit relations among the entire ATLAS$^{3D}$ X-ray sample are $L_{X,GAS} \sim L_K^3$ and $L_{X,GAS} - T_{GAS}^5$, but the rms deviations are still large in both relations (0.7-0.8 dex or a factor of 5-6).
(2) Among normal (non cD) core galaxies which are considered as passively evolving (no star formation), σ—supported (slow rotation) Es, the best fit relations are $L_{X,GAS} \sim L_K^{2.8}$ and $L_{X,GAS} - T_{GAS}^{4.5}$. The most significant result is the tight correlation between $L_{X,GAS}$ and $T_{GAS}$. The rms deviation is 0.2 dex. If we exclude 3 galaxies with indirect signatures of recent star formation, the rms deviation of 11 genuine Es is further reduced to 0.13 dex (or a factor of 1.3).
(3) For the gas-rich galaxies ($L_X > 10^{40}$ erg s$^{-1}$), the tight $L_{X,GAS} - T_{GAS}$ relation is consistent with the recent simulations of Negri et al (2014). However, the tight relation unexpectedly extends to gas-poor galaxies ($L_{X,GAS}$ = a few x $10^{38}$ - $10^{40}$ erg s$^{-1}$), where the gas is expected to be in a wind/outflow state. Our result may suggest the presence of small bound hot halos even in this low luminosity range.
(4) We found that cD-type galaxies (central dominant galaxies in groups and clusters) follow a similar scaling relations, but have an order of magnitude higher $L_{X,GAS}$ than giant elliptical galaxies with similar $L_K$ and $T_{GAS}$. Based on the presence of extended cold gas in several cDs, we suggest that enhanced cooling in cDs, which have higher hot gas densities and lower entropies, could lower $T_{GAS}$ to the range observed in giant Es.
(5) Among coreless ETGs, $L_{X,GAS}$ and $T_{GAS}$ do not correlate at all, similar to what is found among spiral galaxies, suggesting that various additional mechanisms (e.g., rotation, flattening, star formation) play a significant role to disrupt the hot ISM.

We thank Silvia Pellegrini for helpful discussions. The data analysis was supported by the CXC CIAO software and CALDB. We have used the NASA NED and ADS facilities, and have extracted archival data from the *Chandra* Data Archive. This work was supported by the *Chandra* GO grant AR4-15005X, by 2014 Smithsonian Competitive Grant Program for Science and by NASA contract NAS8-03060 (CXC).

Table 1  Sample early type galaxies and their properties

| name | T | d (Mpc) | $M_K$ (mag) | $\lambda_R$ | e | rot | $\sigma_o$ (km s$^{-1}$) | core | log($M_{H2}$) Mo | log($M_{HI}$) Mo | log($M_{HIc}$) Mo | dust | age (Gyr) | ref |
|---|---|---|---|---|---|---|---|---|---|---|---|---|---|---|
| (1) | (2) | (3) | (4) | (5) | (6) | (7) | (8) | (9) | (10) | (11) | (12) | (13) | (14) | (15) |
| N0474 | -2.0 | 30.9 | -23.91 | 0.213 | 0.192 | F | 167.1 | 2 | < 7.68 | | | N | 7.1 | 6 |
| N0524 | -1.2 | 23.3 | -24.71 | 0.325 | 0.034 | F | 243.2 | 1 | 7.97 | | | N | 12.2 | 6 |
| N0821 | -4.8 | 23.4 | -23.99 | 0.273 | 0.392 | F | 199.5 | 3 | < 7.52 | < 6.91 | < 6.53 | N | 8.9 | 1 |
| N1023 | -2.7 | 11.1 | -24.01 | 0.391 | 0.363 | F | 207.0 | 3 | < 6.79 | 9.29 | 6.84 | N | 4.7 | 5 |
| N1266 | -2.1 | 29.9 | -22.93 | 0.638 | 0.193 | F | 78.9 | 9 | 9.28 | | | F | | |
| N2768 | -4.4 | 21.8 | -24.71 | 0.253 | 0.472 | F | 202.3 | 2 | 7.64 | 7.81 | < 6.61 | N | 10.0 | 3 |
| N2778 | -4.8 | 22.3 | -22.23 | 0.572 | 0.224 | F | 164.1 | 3 | < 7.48 | < 7.06 | < 6.68 | N | 6.4 | 1 |
| N3245 | -2.1 | 20.3 | -23.69 | 0.626 | 0.442 | F | 209.9 | 3 | 7.27 | < 7.00 | < 6.61 | N | 5.8 | 8 |
| N3377 | -4.8 | 10.9 | -22.76 | 0.522 | 0.503 | F | 145.5 | 3 | < 6.96 | < 6.52 | < 6.14 | N | 3.6 | 1 |
| N3379 | -4.8 | 10.3 | -23.80 | 0.157 | 0.104 | F | 213.3 | 1 | < 6.72 | < 6.49 | < 6.11 | N | 10.0 | 1 |
| N3384 | -2.7 | 11.3 | -23.52 | 0.407 | 0.065 | F | 161.4 | 3 | < 7.11 | 7.25 | < 6.19 | N | 3.2 | 5 |
| N3412 | -2.0 | 11.0 | -22.55 | 0.406 | 0.441 | F | 105.2 | 3 | < 6.96 | < 6.55 | < 6.17 | N | 1.9 | 2 |
| N3414 | -2.0 | 24.5 | -23.98 | 0.073 | 0.194 | S | 231.7 | 3 | < 7.19 | 8.28 | < 7.70 | N | 3.6 | 5 |
| N3599 | -2.0 | 19.8 | -22.22 | 0.282 | 0.080 | F | 74.8 | 3 | 7.36 | < 7.03 | < 6.64 | N | 1.2 | 8 |
| N3607 | -3.1 | 22.2 | -24.74 | 0.209 | 0.185 | F | 229.1 | 9 | 8.42 | < 6.92 | < 6.53 | D | 3.6 | 2 |
| N3608 | -4.8 | 22.3 | -23.65 | 0.043 | 0.190 | S | 193.2 | 1 | < 7.58 | 7.16 | < 6.53 | N | 8.0 | 1 |
| N3665 | -2.1 | 33.1 | -24.92 | 0.410 | 0.216 | F | 224.9 | 9 | 8.91 | < 7.43 | < 7.05 | D | 2.0 | 8 |
| N3945 | -1.2 | 23.2 | -24.31 | 0.524 | 0.090 | F | 182.8 | 3 | < 7.50 | 8.85 | < 6.73 | FBR | | |
| N3998 | -2.1 | 13.7 | -23.33 | 0.454 | 0.170 | F | 278.6 | 2 | < 7.06 | 8.45 | 7.42 | N | | |
| N4026 | -1.8 | 13.2 | -23.03 | 0.548 | 0.556 | F | 190.1 | 3 | < 6.99 | 8.50 | < 7.14 | N | | |
| N4036 | -2.6 | 24.6 | -24.40 | 0.685 | 0.555 | F | 188.8 | 9 | 8.13 | 8.41 | < 6.80 | F | | |
| N4111 | -1.4 | 14.6 | -23.27 | 0.619 | 0.582 | F | 158.9 | 9 | 7.22 | 8.81 | 6.94 | N | | |
| N4203 | -2.7 | 14.7 | -23.44 | 0.305 | 0.154 | F | 159.2 | 3 | 7.39 | 9.15 | 7.03 | N | | |
| N4233 | -2.0 | 33.9 | -23.88 | 0.564 | 0.277 | F | 217.3 | 9 | < 7.89 | | | F | | |
| N4251 | -1.9 | 19.1 | -23.68 | 0.584 | 0.508 | F | 137.1 | 9 | < 7.11 | < 6.97 | < 6.58 | N | 1.9 | 2 |
| N4261 | -4.8 | 30.8 | -25.18 | 0.085 | 0.222 | S | 294.4 | 1 | < 7.68 | | | N | 16.3 | 1 |
| N4278 | -4.8 | 15.6 | -23.80 | 0.178 | 0.103 | F | 248.3 | 1 | < 7.45 | 8.80 | 6.06 | N | 12.0 | 1 |
| N4342 | -3.4 | 16.5 | -22.07 | 0.528 | 0.442 | F | 252.9 | 3 | < 7.24 | | | N | | |
| N4365 | -4.8 | 23.3 | -25.21 | 0.088 | 0.254 | S | 255.9 | 1 | < 7.62 | | | N | 5.9 | 3 |
| N4374 | -4.3 | 18.5 | -25.12 | 0.024 | 0.147 | S | 288.4 | 1 | < 7.23 | < 7.26 | < 6.88 | N | 12.8 | 1 |
| N4382 | -1.3 | 17.9 | -25.13 | 0.163 | 0.202 | F | 183.7 | 1 | < 7.39 | < 6.97 | < 6.59 | N | 1.6 | 2 |
| N4406 | -4.8 | 16.8 | -25.04 | 0.052 | 0.211 | S | 216.8 | 1 | < 7.40 | 8.00 | < 6.40 | N | | |
| N4459 | -1.4 | 16.1 | -23.89 | 0.438 | 0.148 | F | 178.6 | 3 | 8.24 | < 6.91 | < 6.53 | D | 1.9 | 5 |
| N4472 | -4.8 | 17.1 | -25.78 | 0.077 | 0.172 | F | 288.4 | 1 | < 7.25 | | | N | 9.6 | 1 |
| N4473 | -4.7 | 15.3 | -23.77 | 0.229 | 0.421 | F | 193.6 | 1 | < 7.07 | < 6.86 | < 6.47 | N | 4.0 | 3 |
| N4477 | -1.9 | 16.5 | -23.75 | 0.221 | 0.135 | F | 171.4 | 2 | 7.54 | < 6.95 | < 6.56 | N | 11.7 | 6 |
| N4486 | -4.3 | 17.2 | -25.38 | 0.021 | 0.037 | S | 314.1 | 1 | < 7.17 | | | N | 19.6 | 3 |
| N4494 | -4.8 | 16.6 | -24.11 | 0.212 | 0.173 | F | 154.2 | 3 | < 7.25 | < 6.84 | < 6.46 | N | 6.7 | 8 |
| N4526 | -1.9 | 16.4 | -24.62 | 0.453 | 0.361 | F | 235.0 | 9 | 8.59 | | | D | 1.6 | 4 |
| N4552 | -4.6 | 15.8 | -24.29 | 0.049 | 0.047 | S | 261.8 | 1 | < 7.28 | < 6.87 | < 6.48 | N | 12.4 | 1 |
| N4564 | -4.8 | 15.8 | -23.08 | 0.619 | 0.560 | F | 174.2 | 3 | < 7.25 | < 6.91 | < 6.53 | N | 5.9 | 2 |
| N4596 | -0.9 | 16.5 | -23.63 | 0.297 | 0.254 | F | 151.0 | 3 | 7.31 | < 7.13 | < 7.13 | N | | |
| N4621 | -4.8 | 14.9 | -24.14 | 0.291 | 0.365 | F | 223.9 | 3 | < 7.13 | < 6.86 | < 6.48 | N | 15.8 | 3 |
| N4636 | -4.8 | 14.3 | -24.36 | 0.036 | 0.094 | S | 199.5 | 1 | < 6.87 | | | N | 13.5 | 7 |
| N4649 | -4.6 | 17.3 | -25.46 | 0.127 | 0.156 | F | 314.8 | 1 | < 7.44 | < 7.18 | < 7.19 | N | 14.1 | 1 |
| N4697 | -4.4 | 11.4 | -23.93 | 0.322 | 0.447 | F | 180.7 | 3 | < 6.86 | | | N | 8.3 | 1 |
| N4710 | -0.9 | 16.5 | -23.53 | 0.652 | 0.699 | F | 105.0 | 9 | 8.72 | 6.84 | 6.71 | D | | |
| N5198 | -4.8 | 39.6 | -24.10 | 0.061 | 0.146 | S | 201.4 | 1 | < 7.89 | 8.49 | < 6.98 | N | 15.1 | 5 |
| N5322 | -4.8 | 30.3 | -25.26 | 0.073 | 0.307 | S | 248.3 | 1 | < 7.76 | < 7.34 | < 6.96 | N | 2.4 | 8 |
| N5422 | -1.5 | 30.8 | -23.69 | 0.600 | 0.604 | F | 174.6 | 2 | < 7.78 | 7.87 | 7.43 | D | | |
| N5576 | -4.8 | 24.8 | -24.15 | 0.102 | 0.306 | S | 185.8 | 2 | < 7.60 | | | N | 2.5 | 3 |
| N5582 | -4.9 | 27.7 | -23.28 | 0.564 | 0.320 | F | 151.4 | 9 | < 7.67 | 9.65 | < 6.88 | N | 17.0 | 2 |
| N5638 | -4.8 | 25.6 | -23.80 | 0.265 | 0.091 | F | 159.2 | 9 | < 7.60 | | | N | 9.5 | 1 |
| N5813 | -4.8 | 31.3 | -25.09 | 0.071 | 0.170 | S | 225.9 | 1 | < 7.69 | | | N | 16.6 | 1 |
| N5838 | -2.6 | 21.8 | -24.13 | 0.521 | 0.361 | F | 276.1 | 3 | < 7.56 | | | N | 11.2 | 6 |
| N5845 | -4.9 | 25.2 | -22.92 | 0.404 | 0.264 | F | 269.8 | 3 | < 7.50 | | | N | 8.9 | 5 |
| N5846 | -4.7 | 24.2 | -25.01 | 0.032 | 0.062 | S | 231.7 | 1 | < 7.78 | | | N | 14.2 | 1 |
| N5866 | -1.3 | 14.9 | -24.00 | 0.319 | 0.566 | F | 160.3 | 9 | 8.47 | 6.96 | 6.67 | D | 1.8 | 2 |
| N6017 | -5.0 | 29.0 | -22.52 | 0.429 | 0.455 | F | 109.9 | 9 | < 7.73 | | | D | | |
| N6278 | -1.9 | 42.9 | -24.19 | 0.576 | 0.409 | F | 217.8 | 3 | < 7.98 | < 7.67 | < 7.28 | N | | |
| N7457 | -2.6 | 12.9 | -22.38 | 0.519 | 0.470 | F | 74.8 | 3 | < 6.96 | < 6.61 | < 6.22 | N | 3.4 | 6 |

Note.
1. galaxy NGC name
2. morphological type from Cappellari, M. et al. (2011 MNRAS 413, 813)
3. distance mainly from Tonry et al. (2001, ApJ, 546, 681) and from Cappellari, M. et al. (2011 MNRAS 413, 813) if not listed in the former
4. absolute K-band mag from 2MASS
5. $\lambda_R = \langle RV \rangle / [R(V^2+\sigma^2)^{0.5}]$ measured at $R_e$ from Emsellem, E., et al. (2011 MNRAS 414, 888)
6. ellipticity measured at $R_e$ from Emsellem, E., et al. (2011 MNRAS 414, 888)
7. F(fast) or S(slow) rotator from Emsellem, E., et al. (2011 MNRAS 414, 2923)
8. central velocity dispersion in $r < 1/8\ R_e$ from Cappellari, M. et al. (2013 MNRAS 432, 1862)
9. central radial profile 1 (core), 2 (intermediate), 3 (cusp), 9 (unknown) from Krajnovic, D. et al. (2013 MNRAS 433, 2812)
10. Mass of $H_2$ molecular gas from Young, L. M. et al. (2011 MNRAS 414, 940)
11. Mass of HI gas from Serra, P. et al. (2012 MNRAS 422, 1835)
12. Mass of central HI gas from Young, L. M. et al. (2014, MNRAS, 444, 3408)
13. dust features: D (dusty disc), F (dusty filament), B (blue nucleus) and BR (blue ring) from Krajnović, D. et al. (2011 MNRAS 414, 2923)
14. ages and references
    1. Thomas et al. 2005, ApJ, 621, 673
    2. Terlevich and Forbes 2002, MNRAS, 330, 547
    3. Howell 2005 AJ, 130, 2065
    4. Gallagher et al. 2008, ApJ, 685, 752
    5. McDermid et al. 2006 MNRAS 373 906
    6. Kuntschner et al. 2010, MNRAS, 408, 97
    7. Annibali et al. 2010, AA, 519, A40
    8. Denicolo et al. 2005, MNRAS, 356, 1440

Table 2  X-ray properties of sample early type galaxies

```
------------------------------------------------------------------------------------------
  name   obsid   r(")   T_Gas(keV)              L_X,Tot   L_X,GAS (10^40 erg s^-1)+
------------------------------------------------------------------------------------------
  N0474  07144    60    0.19  ( 0.00  -  9.00 )  0.831    0.166  ( 0.000   -   0.332 )
  N0524  06778    60    0.50  ( 0.43  -  0.57 )  1.612    1.174  ( 1.027   -   1.325 )
  N0821  multi    30    0.09  ( 0.00  -  9.00 )  0.883    0.025  ( 0.015   -   0.037 )
  N1023  multi    60    0.30  ( 0.29  -  0.32 )  0.615    0.061  ( 0.057   -   0.066 )
  N1266  11578    30    0.81  ( 0.71  -  1.02 )  2.256    0.522  ( 0.459   -   0.632 )
  N2768  09528    60    0.31  ( 0.30  -  0.32 )  2.612    0.753  ( 0.716   -   0.790 )
  N2778  11777    30    0.54  ( 0.19  -  0.99 )  0.296    0.025  ( 0.010   -   0.049 )
  N3245  02926    90    0.30  ( 0.24  -  0.36 )  1.288    0.280  ( 0.211   -   0.349 )
  N3377  02934    30    0.19  ( 0.00  -  0.26 )  0.306    0.010  ( 0.007   -   0.013 )
  N3379  multi    90    0.25  ( 0.23  -  0.27 )  0.864    0.041  ( 0.035   -   0.048 )
  N3384  04692    60    0.26  ( 0.00  -  0.58 )  0.617    0.018  ( 0.004   -   0.033 )
  N3412  04693    30    0.30  ( 0.00  -  9.00 )  0.137    0.025  ( 0.000   -   0.056 )
  N3414  06779    30    0.57  ( 0.37  -  0.74 )  4.477    0.157  ( 0.106   -   0.207 )
  N3599  09556    60    0.16  ( 0.00  -  0.25 )  0.588    0.040  ( 0.014   -   0.068 )
  N3607  02073   240    0.59  ( 0.52  -  0.66 )  2.953    1.600  ( 1.416   -   1.784 )
  N3608  02073    90    0.40  ( 0.34  -  0.49 )  1.147    0.414  ( 0.348   -   0.480 )
  N3665  03222    60    0.34  ( 0.31  -  0.41 )  3.044    1.260  ( 1.036   -   1.482 )
  N3945  06780    30    0.37  ( 0.29  -  0.53 )  1.873    0.300  ( 0.228   -   0.371 )
  N3998  06781    30    0.26  ( 0.22  -  0.32 ) 32.555    0.999  ( 0.761   -   1.236 )
  N4026  06782    30    0.28  ( 0.17  -  1.06 )  0.215    0.023  ( 0.010   -   0.039 )
  N4036  06783    30    0.46  ( 0.32  -  0.61 )  2.998    0.428  ( 0.369   -   0.513 )
  N4111  01578    30    0.47  ( 0.41  -  0.51 )  1.108    0.480  ( 0.446   -   0.515 )
  N4203  10535    60    0.25  ( 0.18  -  0.33 )  6.019    0.208  ( 0.105   -   0.210 )
  N4233  06784    30    0.20     fixed           4.383    0.018  ( 0.003   -   0.107 )
  N4251  04695    60    0.28  ( 0.05  -  0.83 )  0.554    0.048  ( 0.015   -   0.090 )
  N4261  09569    60    0.76  ( 0.75  -  0.77 ) 19.540    6.844  ( 6.735   -   6.952 )
  N4278  multi    90    0.30  ( 0.29  -  0.31 )  3.956    0.245  ( 0.232   -   0.258 )
  N4342  04687    60    0.59  ( 0.55  -  0.64 )  0.427    0.164  ( 0.151   -   0.178 )
  N4365  multi   120    0.46  ( 0.44  -  0.49 )  3.787    0.470  ( 0.451   -   0.490 )
  N4374  00803    -     0.73  ( 0.73  -  0.74 )  9.001    6.650  ( 5.474   -   7.826 )*
  N4382  02016    90    0.39  ( 0.37  -  0.40 )  2.806    0.979  ( 0.944   -   1.014 )
  N4406  00318    -     0.82  ( 0.81  -  0.83 )130.057  127.773  (126.631  - 128.915 )*
  N4459  02927    60    0.40  ( 0.34  -  0.55 )  1.318    0.251  ( 0.204   -   0.292 )
  N4472  00321    -     0.95  ( 0.94  -  0.96 ) 28.180   24.223  ( 22.244  -  26.202 )*
  N4473  04688    60    0.31  ( 0.28  -  0.35 )  0.666    0.126  ( 0.105   -   0.147 )
  N4477  09527    90    0.33  ( 0.32  -  0.34 )  1.556    0.926  ( 0.887   -   0.964 )
  N4486  multi    -     1.50  ( 1.50  -  1.50 )908.124  905.499  (904.186  - 906.812 )*
  N4494  02079   120    0.34  ( 0.25  -  0.86 )  1.546    0.130  ( 0.053   -   0.216 )
  N4526  03925    30    0.31  ( 0.29  -  0.32 )  1.525    0.281  ( 0.262   -   0.299 )
  N4552  02072    90    0.59  ( 0.59  -  0.60 )  4.825    2.154  ( 2.067   -   2.133 )
  N4564  04008    30    0.27  ( 0.00  -  0.72 )  0.285    0.038  ( 0.024   -   0.053 )
  N4596  02928    60    0.21  ( 0.19  -  0.24 )  0.784    0.263  ( 0.223   -   0.303 )
  N4621  02068    90    0.27  ( 0.21  -  0.35 )  1.274    0.063  ( 0.036   -   0.090 )
  N4636  00323    -     0.73  ( 0.72  -  0.73 ) 32.989   31.744  ( 31.121  -  32.367 )*
  N4649  multi    -     0.86  ( 0.86  -  0.86 ) 21.295   18.215  ( 16.675  -  19.755 )*
  N4697  multi   120    0.31  ( 0.31  -  0.32 )  1.160    0.189  ( 0.180   -   0.199 )
  N4710  09512    60    0.33  ( 0.30  -  0.39 )  0.279    0.115  ( 0.098   -   0.133 )
  N5198  06786    30    0.36  ( 0.26  -  0.62 )  1.561    0.281  ( 0.160   -   0.396 )
  N5322  06787    30    0.33  ( 0.30  -  0.38 )  1.576    0.665  ( 0.567   -   0.761 )
  N5422  multi    30    0.24  ( 0.02  -  0.42 )  0.594    0.039  ( 0.013   -   0.070 )
  N5576  11781    30    0.52  ( 0.30  -  0.72 )  0.482    0.055  ( 0.032   -   0.080 )
  N5582  11361    30    0.30     fixed           0.298    0.029  ( 0.011   -   0.068 )
  N5638  11313    30    0.30     fixed           0.410    0.000  ( 0.000   -   0.111 )
  N5813  05907   120    0.70  ( 0.69  -  0.70 ) 74.016   70.654  ( 70.313  -  70.980 )
  N5838  06788    60    0.42  ( 0.35  -  0.50 )  1.526    0.735  ( 0.623   -   0.844 )
  N5845  04009    30    0.39  ( 0.25  -  0.66 )  0.424    0.056  ( 0.033   -   0.084 )
  N5846  00788    -     0.72  ( 0.72  -  0.73 ) 52.722   50.513  ( 49.407  -  51.618 )*
  N5866  02879    90    0.32  ( 0.30  -  0.34 )  0.891    0.251  ( 0.231   -   0.272 )
  N6017  11363    30    0.30     fixed           0.223    0.000  ( 0.000   -   0.035 )
  N6278  06789    60    0.52  ( 0.30  -  0.76 )  4.483    0.246  ( 0.151   -   0.365 )
  N7457  04697    60    0.30     fixed           0.114    0.012  ( 0.002   -   0.034 )
------------------------------------------------------------------------------------------
```

+ $L_{X,GAS}$ in 0.3-8keV  
\* used ROSAT results from O'Sullivan et al. (2001)